\def\fr#1#2{\hbox{${#1\over #2}$}}

\def\+{{(+)}}  \def\-{ {(-)} }   \def\0{ {(0)} }
\def\1{ {(1)} }  \def\2{ {(2)} }
\def\pd{\partial}

\def\con{\omega}

\def\sq{Q\kern-6pt/}
\def\sQ{Q\kern-12pt\nearrow}
\documentclass[11pt,eqs]{article}
\usepackage{latexsym,graphicx}

\def\be{\begin{equation}}             \def\ee{\end{equation}}
\def\ba{\begin{array}{rcl}}           \def\ea{\end{array}}
\def\beqa{\begin{eqnarray} }          \def\eeqa{\end{eqnarray} }
\def\beqalign{\begin{eqalign}}        \def\eeqalign{\end{eqalign}}
\def\leq#1{\label{eq:#1}}             \def\eq#1{(\ref{eq:#1})}
\def\bsubeq{\begin{subequations}}     \def\esubeq{\end{subequations}}
\def\bitem{\begin{itemize}}           \def\eitem{\end{itemize}}

\def\DJ{\leavevmode\setbox0=\hbox{D}\kern0pt
 \rlap{\kern.04em\raise.188\ht0\hbox{-}}D}
\def\dj{\leavevmode\setbox0=\hbox{d}\kern0pt
 \rlap{\kern.215em\raise.46\ht0\hbox{-}}d}

\newcommand{\bd}{\begin{displaymath}}
\newcommand{\ed}{\end{displaymath}}

\begin{document}
\title{ The geometrical form for the string space-time action
\thanks{Work supported in part by the Serbian Ministry of Science and
Environmental Protection, under contract No. 141036.}}
\author{D. S. Popovi\'c \thanks{e-mail address:popovic@phy.bg.ac.yu} and
B. Sazdovi\'c \thanks{e-mail address:sazdovic@phy.bg.ac.yu}\\
{\it Institute of Physics, 11001 Belgrade, P.O.Box 57, Serbia}}
\date{}
\maketitle

\begin{abstract}
In the present article, we derive the space-time action of the bosonic string
in terms of geometrical quantities. First, we study the space-time geometry
felt by probe bosonic string moving in antisymmetric and dilaton background
fields. We show that the presence of the antisymmetric field leads to the
space-time torsion, and the presence of the dilaton field leads to the
space-time nonmetricity. Using these results we obtain the integration measure
for space-time with stringy nonmetricity, requiring its preservation under
parallel transport. We derive the Lagrangian depending on stringy curvature,
torsion and nonmetricity.
\end{abstract}

\section{Introduction}
\setcounter{equation}{0}

The general relativity is described in terms of torsion free and metric
compatible connection. There are many generalization of this theory which
include nontrivial contribution of torsion and nonmetricity \cite{MBH}.

We are interesting in the theory of gravity obtained from string theory,
describing the massless states of the closed bosonic string. Beside the metric
tensor $G_{\mu \nu}$, it contains the antisymmetric tensor $B_{\mu \nu}$ and
the dilaton field $\Phi$. The space-time field equations of this theory can be
derived from the requirement of Weyl invariance of the quantum world-sheet
theory, as a condition of consistent string theory \cite{CFMP}. It is
nontrivial fact that these field equations can be obtained from a single
space-time action. Consequently, the quantum conformal invariance of the
world-sheet leads to the generalized space-time Einstein equations and
corresponding action. The question is whether there exists the geometrical
interpretation of this action? In our interpretation it means the existence of
generalized connection, so that the above action can be written in terms of
corresponding generalized curvature, torsion and nonmetricity.

There was many attempts to achieve this goal, expressing this action in terms
of geometrical quantities. In ref.\cite{DT} there is a restriction on the
space-time dimensions (D=2 and D=4), in order to use Hodge dual map. Some
articles (first reference \cite{DT} and \cite{CD,Sa}) investigate
Riemann-Cartan metric compatible space-time, while in second reference
\cite{DT} the space-time is torsion free but with metric non-compatible
connection. The third article \cite{DT}  considers space-time, with nontrivial
both torsion and nonmetricity. The authors assume the form of the torsion and
nonmetricity and restrict considerations on D=4 space-time dimensions.
The torsion is usually connected with field strength of antisymmetric field
\cite{SS}, while in some papers \cite{CD,Sa} the trace of the torsion is
related with gradient of the dilaton field.

In this article we first derive the form of the connection from the
world-sheet equations of motion. It corresponds to new covariant derivative,
which makes easier to perform calculations, including those in quantization
procedure. We find that the string sees the space-time not as a Riemann one,
but as some particular form of the affine space-time, which beside curvature
also depends on torsion and nonmetricity. The features of this geometry define
effective general relativity in the target space.

In Sec. 2, we formulate the theory and shortly repeat some results of ref.
\cite{BS1}.

Starting with the known rules of the space-time parallel transport, in Sec. 3
we introduce the torsion and nonmetricity. We decompose the arbitrary
connection in terms of the Christoffel one, contortion and nonmetricity. With
the help of equations of motion, we find particular form of stringy torsion
and stringy nonmetricity \cite{BS1}. To the space-time felt by the probe
string we will refer as stringy space-time.

In sec.4, we derive the form of the space-time action. We obtain the
integration measure for spaces with nonmetricity from the requirements that
the measure is preserved under parallel transport and that it enables
integration by parts. Our integration measure is a volume-form compatible with
affine connection ref.\cite{Saa}. In particular application, to improve the
standard measure, ref.\cite{Sa} usees the torsion while we use the
nonmetricity. We construct the Lagrangian linear in stringy invariants: scalar
curvature, square of torsion and square of nonmetricity. We discuss the
relation of the space-time action of the present paper with the space-time
action of the papers \cite{CFMP}.

Appendix A is devoted to the world-sheet geometry.

\section{Canonical derivation of field equations}
\setcounter{equation}{0}

The closed bosonic string, propagated in arbitrary background is described by
the sigma model (see \cite{CFMP} and \cite{GSW})
\be
S= \kappa  \int_\Sigma d^2 \xi \sqrt{-g} \left\{ \left[ {1 \over
2}g^{\alpha\beta}G_{\mu\nu}(x) +{\varepsilon^{\alpha\beta} \over \sqrt{-g}}
B_{\mu\nu}(x)\right] \partial_\alpha x^\mu \partial_\beta x^\nu + \Phi(x)
{R}^{(2)} \right\}  \, ,   \leq{ac}
\ee
with $x^\mu$-dependent background fields: metric $G_{\mu \nu}$, antisymmetric
tensor field $B_{\mu\nu}=- B_{\nu\mu}$ and dilaton field $\Phi$. Here,
$g_{\alpha \beta}$ is the intrinsic world-sheet metric and ${R}^{(2)}$ is
corresponding scalar curvature. Let $x^\mu(\xi) \,  (\mu =0,1,...,D-1)$ be the
coordinates of the $D$ dimensional space-time $M_D$ and $\xi^\alpha \, (\xi^0
=\tau , \xi^1=\sigma)$ the coordinates of two dimensional world-sheet
$\Sigma$, spanned by the string. The corresponding derivatives we will denote
as $\partial_\mu \equiv {\partial \over \partial x^\mu}$ and $\partial_\alpha
\equiv {\partial \over
\partial \xi^\alpha}$.

Let us briefly review the canonical analysis and derivation of the field
equations obtained in ref. \cite{BS1}. Restricting consideration to the
condition $a^2 \equiv G^{\mu \nu} a_\mu a_\nu  \neq 0$ ($a_\mu =
\partial_\mu \Phi$) we define the currents
\be
J_{\pm \mu} =P^T{}_\mu{}^\nu j_{\pm \nu} +{a_\mu \over 2
a^2}i_\pm^\Phi = j_{\pm \mu} -{ a_\mu \over a^2} j \,  ,  \leq{J}
\ee
\be
i_\pm^F= {a^\mu \over a^2} j_{\pm \mu}-{1 \over 2 a^2} i_\pm^\Phi
\pm 2 \kappa {F^\prime } \,  , \qquad i_\pm^\Phi= \pi_F \pm 2\kappa
\Phi' \,  ,   \leq{2.8}
\ee
where
\be
j_{\pm \mu} =\pi_\mu +2\kappa \Pi_{\pm \mu \nu} {x^\nu}' \,  ,
\qquad  \Pi_{\pm \mu \nu} \equiv  B_{\mu \nu} \pm {1 \over 2}
G_{\mu \nu} \,  , \leq{jmi}
\ee
and
\be
j=a^\mu j_{\pm \mu} -{1 \over 2} i_\pm^\Phi =a^2 (i_\pm^F \mp 2 \kappa F^\prime)  \,   .   \leq{j}
\ee
Here $\pi_\mu$ and $\pi_F$ are canonically conjugate momenta to the variables
$x^\mu$ and $F$.

Up to boundary term, the canonical Hamiltonian density has the standard form
\be
{\cal H}_c= h^- T_- + h^+ T_+    \,   ,  \leq{hc}
\ee
with the energy momentum tensor components
\be
T_\pm =\mp {1 \over 4\kappa} \left(G^{\mu\nu} J_{\pm \mu} J_{\pm \nu} +
i_\pm^F i_\pm^\Phi \right)
+{1 \over 2} i_\pm^{\Phi \prime} =
\mp {1 \over 4\kappa} \left( G^{\mu\nu}
j_{\pm \mu} j_{\pm \nu} -{j^2 \over a^2} \right) +
{1 \over 2} ({i_\pm^\Phi}' - F' i_\pm^{\Phi})
 \,   .  \leq{emt}
\ee

In spite of their complicated expressions, the same chirality energy-momentum
tensor components satisfy two independent copies of Virasoro algebras,
\be
\{ T_\pm(\sigma) , T_\pm (\bar \sigma) \}= -[ T_\pm(\sigma) +T_\pm({\bar \sigma}) ]
{ \delta^\prime}(\sigma - \bar \sigma) \,   ,   \leq{Vir}
\ee
while the opposite chirality components commute $\{ T_\pm , T_\mp \}= 0$.

\subsection{Equations of motion}

In ref.\cite{BS1}, using canonical approach, we derived the following
equations of motion
\be
[J^\mu]  \equiv  {\nabla}_\mp \partial_\pm  x^\mu + {}^\star
\Gamma_{\mp \rho \sigma}^\mu
\partial_\pm x^\rho \partial_\mp x^\sigma =0  \,  ,    \leq{lJ}
\ee
\be
[h^\pm]  \equiv  G_{\mu \nu}  \partial_\pm  x^\mu \partial_\pm x^\nu -2
{\nabla}_\pm \partial_\pm \Phi =0   \,  ,  \leq{lh}
\ee
\be
[i^F]   \equiv  {R}^{(2)} + {2 \over a^2} (D_{\mp \mu} a_\nu)
\partial_\pm  x^\nu \partial_\mp  x^\mu =0   \,  ,     \leq{lF}
\ee
where the variables in the parenthesis denote the currents corresponding
to this equation. The expression
\be
{}^\star \Gamma^\rho_{\pm \nu \mu}= \Gamma^\rho_{\pm \nu \mu} +{a^\rho \over
a^2} D_{\pm \mu} a_\nu  = \Gamma^\rho_{\nu \mu} \pm P^{T \rho}{}_\sigma
B^\sigma_{\nu \mu} +{a^\rho \over a^2} D_{\mu} a_\nu   \,  ,    \leq{cdc}
\ee
which appears in the $[J^\mu]$ equation is a generalized connection, which
full geometrical interpretation we are going to investigate. Under space-time
general coordinate transformations the expression ${}^\star \Gamma^\rho_{\pm
\nu \mu}$ transforms as a connection.

The covariant derivatives with respect to the Christoffel connection
$\Gamma^\rho_{\nu \mu}$ and to the connection $\Gamma^\rho_{\pm \nu \mu}=
\Gamma^\rho_{\nu \mu} \pm B^\rho_{\nu \mu}$, we respectively denote as $D_\mu$
and $D_{\pm \mu}$, while
\be
B_{\mu \nu \rho}= \partial_\mu B_{\nu \rho} + \partial_\nu B_{\rho \mu} +
\partial_\rho B_{\mu \nu}= D_\mu B_{\nu \rho} + D_\nu B_{\rho \mu} + D_\rho
B_{\mu \nu}   \,  ,       \leq{fsB}
\ee
is the field strength of the antisymmetric tensor. The projection operator
which appears in eq. \eq{cdc}
\be
P^T{}_{\mu \nu} = G_{\mu \nu} - {a_\mu a_\nu \over a^2}  \equiv G_{\mu
\nu}^{D-1} \,  ,   \leq{PT}
\ee
is the induced metric on the $D-1$ dimensional submanifold defined by the
condition $\Phi(x)= const$.

In  \eq{lJ} and \eq{lF} we omit the
currents $\pm$ indices, because $[J^\mu_+] = [J^\mu_-]$ and $[i^F_+] =
[i^F_-]$ as a consequence of the symmetry relations ${}^\star \Gamma_{\mp \rho
\sigma}^\mu = {}^\star \Gamma_{\pm \sigma \rho}^\mu$ and $D_{\mp \mu} a_\nu =
D_{\pm \nu} a_\mu$.

\section{The geometry of space-time seen by the probe string}
\setcounter{equation}{0}

In this section we introduce affine linear connection, torsion and
nonmetricity ( see ref. \cite{MBH} for more details). With the help of string
field equations we derive expressions for stringy connection, torsion and
nonmetricity, recognized by the probe string.

\subsection{Geometry of space-time with torsion and nonmetricity}

In the curved spaces, the operations on tensors are covariant only if they are
realized in the same point. In order to compare the vectors from different
points we need the rule for parallel transport. The parallel transport of the
vector $V^\mu (x)$, from the point $x$ to the point $x+dx$, produce the vector
${}^\circ V^\mu_\parallel = V^\mu + {}^\circ \delta V^\mu$, where
\be
{}^\circ \delta V^\mu = - {}^\circ \Gamma_{\rho \sigma}^\mu V^\rho d x^\sigma  \,  .  \leq{ptcn}
\ee
The variable ${}^\circ \Gamma_{\rho \sigma}^\mu$ is the {\bf affine linear
connection}. The covariant derivative is define in the standard form
\be
{}^\circ D V^\mu = V^\mu (x+dx) - {}^\circ V^\mu_\parallel = d V^\mu -
{}^\circ \delta V^\mu = (\partial_\nu V^\mu + {}^\circ \Gamma_{\rho \nu}^\mu
V^\rho ) d x^\nu \equiv  {}^\circ D_\nu V^\mu d x^\nu     \,   .
\ee

The antisymmetric part of the affine connection is the {\bf torsion}
\be
{}^\circ T^\rho_{\mu \nu} ={}^\circ \Gamma^\rho_{\mu \nu} - {}^\circ \Gamma^\rho_{\nu \mu} \,  .
\ee
It has a simple geometrical interpretation because it measures the non-closure
of the curved "parallelogram".

The {\bf metric tensor} $G_{\mu \nu}$ is independent variable, which enables
calculation of the scalar product $VU= G_{\mu \nu} V^\mu U^\nu$, in order to
measure lengths and angles.

We already learned, that covariant derivative is responsible for the
comparison of the vectors from different points. What variable is responsible
for comparison of the lengths of the vectors? The squares of the lengths of
the vectors: $V^\mu(x)$ and its parallel transport to the point $x+dx$,
${}^\circ V^\mu_\parallel$, are defined respectively as $V^2(x)= G_{\mu
\nu}(x) V^\mu(x) V^\nu (x)$ and ${}^\circ V_\parallel^2(x+dx)= G_{\mu
\nu}(x+dx){}^\circ V^\mu_\parallel \, {}^\circ V^\nu_\parallel$. If we
remember the invariance of the scalar product under the parallel transport,
than the difference of the squares of the vectors is
\be
{}^\circ \delta V^2 =  {}^\circ V_\parallel^2 (x+dx) - V^2 (x)  =
[G_{\mu \nu}(x+dx) - G_{\mu \nu}(x) - {}^\circ \delta G_{\mu \nu}(x)]
\, {}^\circ V^\mu_\parallel \, \, {}^\circ V^\nu_\parallel    \,   .
\ee
Up to the higher order terms we have
\be
{}^\circ \delta V^2 =  [d G_{\mu \nu}(x) - {}^\circ \delta G_{\mu \nu}(x)]
V^\mu V^\nu =
 {}^\circ D G_{\mu \nu} V^\mu V^\nu  \equiv - d x^\rho \, {}^\circ\! Q_{\rho \mu \nu} V^\mu V^\nu \,  , \leq{len}
\ee
where we introduced the {\bf nonmetricity} as a covariant derivative of the
metric tensor
\be
{}^\circ Q_{\mu \rho \sigma}=- {}^\circ D_\mu G_{\rho \sigma}  \,  .  \leq{nm}
\ee
Beside the length, the nonmetricity also changes the angle between the
vectors $V_1^\mu$ and $V_2^\mu$, according to the relation
\be
{}^\circ \delta \cos( \angle ( V_1 , V_2)) =  {-1 \over 2 \sqrt{V_1^2 V_2^2 }}
\left[2 V_1^\rho V_2^\sigma - \left( {V_1^\rho V_1^\sigma \over V_1^2} +
{V_2^\rho V_2^\sigma \over V_2^2} \right) (V_1 V_2) \right] {}^\circ  Q_{\mu
\rho \sigma} d x^\mu       \,  .    \leq{ang}
\ee

Note that we performed the parallel transport of the vectors, but not of the
metric tensor. It means that for the length calculation in the point $x +dx$,
we used the metric tensor $G_{\mu \nu}(x + dx)$, which lives in this point,
and not the tensor $G_{\mu \nu} + {}^\circ \delta G_{\mu \nu}$ obtained after
parallel transport from the point $x$. The requirement for the equality of
these two tensors is known in the literature as a metric postulate. In fact,
it is just compatibility between the metric and connection, such that metric
after parallel transport is equal to the local metric. Here we will not accept
this requirement, because the difference of these two tensors is the origin of
the nonmetricity. So, the nonmetricity measures the deformation of lengths and
angles during the parallel transport.

We also define the Weyl vector as
\be
{}^\circ q_\mu  = {1 \over D} G^{\rho \sigma } {}^\circ Q_{\mu \rho \sigma}
\,    ,     \leq{wv}
\ee
where $D$ is the number of space-time dimensions. When the traceless part of
the nonmetricity vanishes
\be
{}^\circ \sQ_{\mu \rho \sigma} \equiv {}^\circ Q_{\mu \rho \sigma} - G_{\rho
\sigma} {}^\circ q_\mu =0     \,       ,   \leq{ap}
\ee
the parallel transport preserves the angles but not the lengths. Such geometry
is known as a Weyl geometry.

Following  paper \cite{MBH}, we can decompose the connection ${}^\circ
\Gamma^\mu_{\nu \rho}$ in terms of the Christoffel connection, contortion and
nonmetricity. If we introduce the Schouten braces according to the relation
\be
\{ \mu \rho \sigma \} = \sigma \mu \rho + \rho \sigma \mu - \mu \rho \sigma  \,       ,
\ee
then the Christoffel connection, can be expressed as $\Gamma_{\mu , \rho
\sigma} = \frac{1}{2} \partial_{ \{ \mu } G_{ \rho \sigma \} }$. The
contortion ${}^\circ K_{\mu \rho \sigma}$ is defined in terms of the torsion
\be
{}^\circ K_{\mu \rho \sigma} = \frac{1}{2} {}^\circ T_{ \{ \sigma \mu \rho \}
} = \frac{1}{2} ({}^\circ T_{\rho \sigma \mu} + {}^\circ T_{\mu \rho \sigma} -
{}^\circ T_{\sigma \mu \rho})    \,  .  \leq{K}
\ee

The Schouten braces of the nonmetricity can be solved in terms of the
connection, producing
\be
{}^\circ \Gamma_{\mu ,\rho \sigma} = \Gamma_{\mu ,\rho \sigma} + {}^\circ
K_{\mu \rho \sigma} + \frac{1}{2} {}^\circ Q_{\{ \mu \rho \sigma \} }   \,   .
\leq{cde}
\ee
The first term is the Christoffel connection , which depends on the metric but
which does not transforms as a tensor. The second one is the contortion \eq{K}
and the third one is Schouten braces of the nonmetricity \eq{nm}. The last two
terms transform as a tensors.

\subsection{Stringy torsion and nonmetricity}

The manifold $M_D$, together with the affine connection ${}^\circ
\Gamma^\mu_{\nu \rho}$ and the metric $G_{\mu \nu}$, define the affine
space-time $A_D \equiv ( M_D, {}^\circ \Gamma, G)$. To the connection \eq{cdc}
we will refer as  the {\bf stringy connection} and to the corresponding
space-time $S_D \equiv (M_D, {}^\star \Gamma_\pm, G)$, observed by the string
propagating in the background $G_{\mu \nu}$, $B_{\mu \nu}$ and $\Phi$, we will
refer as the {\bf stringy space-time}.

The antisymmetric part of the stringy connection is the {\bf stringy torsion}
\be
{}^\star T_\pm{}^\rho_{\mu \nu} ={}^\star \Gamma_\pm{}^\rho_{\mu \nu} -
{}^\star \Gamma_\pm{}^\rho_{\nu \mu} = \pm 2 P^{T \rho}{}_\sigma
B^\sigma_{\mu \nu}      \,  .   \leq{T}
\ee
It is the transverse projection of the field strength of the antisymmetric
tensor field $B_{\mu \nu}$. The form of the eq.\eq{fsB} suggests that $B_{\mu
\nu}$ is a torsion potential \cite{CZ}.

The presence of the dilaton field $\Phi$ leads to breaking of the space-time
metric postulate. The non-compatibility of  the metric $G_{\mu \nu}$  with the
stringy connection ${}^\star \Gamma^\mu_{\pm \nu \rho}$ is measured by the
{\bf stringy nonmetricity}
\be
{}^\star Q_{\pm \mu \rho \sigma} \equiv  -{}^\star D_{\pm \mu} G_{\rho \sigma} =
{1 \over a^2} D_{\pm \mu} (a_\rho a_\sigma )       \,   .   \leq{mp}
\ee
Consequently, during  stringy parallel transport, the lengths and angles
deformations depend on the vector field $a_\mu$.

The stringy Weyl vector
\be
{}^\star q_\mu  = {1 \over D} G^{\rho \sigma } {}^\star Q_{\pm \mu \rho
\sigma}= {-4 \over D} \partial_\mu \varphi    \,    ,     \leq{Wvs}
\ee
is a gradient of new scalar field $\varphi$, defined by the expression
\be
\varphi=-{1 \over 4} \ln a^2 = -{1 \over 4} \ln(G^{\mu \nu} \partial_\mu \Phi
\partial_\nu \Phi) \,  .   \leq{sf}
\ee

The stringy angle preservation relation
\be
{}^\star \sQ_{\pm \mu \rho \sigma} = {}^\star Q_{\pm \mu \rho \sigma} -
G_{\rho \sigma} {}^\star q_\mu =0     \,       ,   \leq{aps}
\ee
is a condition on the dilaton field $\Phi$. Generally, in stringy geometry both
the lengths and the angles could be changed under the parallel transport.

Using the relation
\be
{}^\star K_{\pm \mu \rho \sigma} + \frac{1}{2} {}^\star Q_{\pm \{ \mu \rho
\sigma \} } = \pm \frac{1}{2} {}^\star T_{\mu \rho \sigma} + \frac{1}{2}
{}^\star Q_{\{ \mu \rho \sigma \} }  \,   ,
\ee
instead \eq{cde}, we can write
\be
{}^\star \Gamma_{\pm \mu ,\rho \sigma} = \Gamma_{\mu ,\rho \sigma} \pm
\frac{1}{2} {}^\star T_{\mu \rho \sigma} + \frac{1}{2} {}^\star Q_{\{ \mu \rho
\sigma \} }   \,   ,     \leq{cde1}
\ee
where the quantities ${}^\star T_{\mu \rho \sigma} = 2  P^T{}_\mu^\nu B_{\nu
\rho \sigma}$ and ${}^\star Q_{ \mu \rho \sigma} = -{}^\star D_{\mu} G_{\rho
\sigma} = {1 \over a^2} D_{\mu} (a_\rho a_\sigma )$ do not depend on $\pm$
indices. In fact, the last term is ${}^\star Q_{\{ \mu \rho \sigma \} } = 2
\frac{a_\mu}{a^2} D_\rho a_\sigma$, so that we can recognize the starting
expression \eq{cdc}.

\section{The space-time action}
\setcounter{equation}{0}

The space-time field equations for background fields, derived as a quantum
consistency condition of string theory \cite{CFMP}, has a form
\begin{eqnarray}
&& \beta^G_{\mu \nu} \equiv  R_{\mu \nu} - \fr{1}{4} B_{\mu \rho
\sigma}
B_{\nu}{}^{\rho \sigma} +2 D_\mu a_\nu =0   \,  , \\
&& \beta^B_{\mu \nu} \equiv  D_\rho B^\rho{}_{\mu \nu} -2 a_\rho
 B^\rho{}_{\mu \nu} = 0   \,  ,   \\
&& \beta^\Phi \equiv 4 \pi \kappa {D-26 \over 3} -
 R + \fr{1}{12} B_{\mu \rho \sigma}
B^{\mu \rho \sigma} - 4 D_\mu a^\mu + 4 a^2 = 0  \,  ,
\end{eqnarray}
so that the  world-sheet theory is Weyl invariant.
Here $R_{\mu \nu}$, $R$ and $D_\mu$ are space-time Ricci tensor, scalar
curvature and covariant derivative, respectively, while $B_{\mu \rho \sigma}$
is field strength of the field $B_{\mu \nu}$ and $a_\mu=\partial_\mu \Phi$.

These field equations can be derived from a single space-time action
\be
S= \int dx \sqrt{-G} e^{-2 \Phi} [ R -{1 \over 12} B^2 +4 (\partial \Phi)^2 ]
\, ,
\ee
where $B^2 = B_{\mu \nu \rho} B^{\mu \nu \rho}$ and $(\partial \Phi)^2 =
G^{\mu \nu}  \partial_\mu \Phi  \partial_\nu \Phi$.

The action is defined up to the total derivative. So, it depends on some
constant parameter $\zeta$
\be
S_\zeta = S + \zeta \int dx  \partial_\mu (\sqrt{-G} G^{\mu \nu} \partial_\nu e^{-2 \Phi} ) \, ,
\ee
and can be rewritten in the form
\be
S_\zeta = \int dx \sqrt{-G} e^{-2 \Phi} [ R -{1 \over 12} B^2 + 4 (1+\zeta)
(\partial \Phi)^2 -2 \zeta D^2 \Phi] \, ,
\ee
where $D^2 \Phi = G^{\mu \nu} D_\mu \partial _\nu \Phi$. For simplicity, in
order to exclude the third term we adopt $\zeta =-1$ and obtain
\be
S_{\zeta=-1} = \int dx \sqrt{-G} e^{-2 \Phi} [ R -{1 \over 12} B^2 + 2 D^2 \Phi]
\equiv \int dx \sqrt{-G} e^{-2 \Phi}  {\cal L} \, .  \leq{la}
\ee

Using the stringy geometry introduced in the previous section, we are going
to reproduce above space-time actions. Generally, it has the form
\be
{}^\star S = \int d^D x \,\, {}^\star \Omega \,\, {}^\star {\cal L}  \,   ,  \leq{sta}
\ee
where ${}^\star \Omega$ is a measure factor, and ${}^\star {\cal L}$ is a
Lagrangian which depends on the space-time field strengths.

\subsection{The space-time measure}

We define the invariant measure, requiring that:

1. It is invariant under space-time general coordinate transformations.

2. It is preserved under parallel transport, which is equivalent to the
condition ${}^\star D_{\pm \mu} {}^\star \Omega = 0$.

3. It enable integration by parts, which can be achieved with help of of the
Leibniz rule and the relation
\be
\int d^D x \, {}^\star \Omega \, {}^\star D_{\pm \mu} V^\mu = \int d^D x
\partial_\mu ({}^\star \Omega V^\mu)  \,   ,  \leq{ibp}
\ee
so that we are able to use Stoke's theorem.

For Riemann and Riemann-Cartan space-times, the solution for the measure
factor is well known $\Omega = \sqrt{-G}$ $(G = \det G_{\mu \nu})$. For spaces
with nonmetricity, this standard measure is not preserved under the parallel
transport, and requirements 2. and 3. are not satisfy. Instead to change the
connection and find volume-preserving one, as has been done in ref.
\cite{MBH}, we prefer to change the measure.

Let us try to find the stringy measure in the form ${}^\star \Omega=
\Lambda(x) \sqrt{-G}$. In order to be preserved under the parallel transport
with the stringy connection, it must satisfy the condition
\be
{}^\star D_{\pm \mu} ( \sqrt{-G} \Lambda ) = \partial_\mu ( \sqrt{-G} \Lambda ) -
{}^\star \Gamma^\rho_{\pm \mu \rho} \sqrt{-G} \Lambda  = 0       \,  .
\ee
Using the relation
\be
{}^\star \Gamma^\rho_{\pm \mu \rho} = \partial_\mu \ln \left( \sqrt{- G }e^{-2
\varphi}\right) = \Gamma^\rho_{\pm \mu \rho} + \frac{D}{2} {}^\star q_\mu
\,  ,  \leq{Grr}
\ee
we find the equation for $\Lambda \,$,  $\,\,\partial_\mu \Lambda =
\frac{D}{2} {}^\star q_\mu \Lambda$. The fact that the stringy Weyl vector
${}^\star q_\mu$ is a gradient of the scalar field $\varphi$, defined in
\eq{sf}, help us to find the solution $\Lambda = e^{-2 \varphi}$. The stringy
measure factor, preserved under parallel transport with the connection
${}^\star \Gamma^\mu_{\pm \nu}$, obtains the form
\be
{}^\star \Omega = \sqrt{-G} e^{-2 \varphi}    \,   .  \leq{smf}
\ee
Consequently, we have ${}^\star \Gamma^\rho_{\pm \mu \rho} =  \partial_\mu \ln
{}^\star \Omega$, and  \eq{ibp} is satisfied. So, if we use the stringy
measure ${}^\star \Omega$ we can  integrated by parts and all requirements are
satisfied.

The above measure is a volume-form compatible with the connection of the
ref.\cite{Saa}. In our case only nonmetricity contributes to the improvement
because in stringy geometry the torsion contribution vanishes, ${}^\star
T_\pm{}^\rho{}_{\mu \rho} =0$.

The measure factor in \eq{la}, $\sqrt{-G} e^{-2 \Phi}$, has the same form as
the one in the present paper and confirms the existence of some space-time
nonmetricity. The requirement of the full measures equality, $\varphi = \Phi$,
leads to the Liouville like equation for the dilaton field
\be
G^{\mu \nu} \partial_\mu \Phi  \partial_\nu \Phi - e^{-4 \Phi} = 0    \,  .  \leq{lle}
\ee
For $D=2$ it turns to the real Liouville equation.

\subsection{The space-time Lagrangian}

We are going to reproduce the Lagrangian defined in \eq{la} with suitable
combinations of the stringy scalar curvature, defined in the standard way with
the stringy connection  \eq{cdc}
\be
{}^\star R_\pm = R - B^2 +2 D^2 \varphi - 4 (\partial \varphi)^2 + e^{4 \varphi}
[2 (aB)^2 + 2 a^\mu \partial_\mu (Da) + a^\mu D_\mu (Da) + (Da)^2] \,  ,
\ee
the stringy torsion \eq{T}
\be
{}^\star T^\rho{}_{\pm \mu \nu} = \pm \left[ 2 B^\rho{}_{\mu \nu} - 2 e^{4 \varphi}
a^\rho (aB)_{\mu \nu} \right]  \,  ,
\ee
and the stringy nonmetricity \eq{mp}
\be
{}^\star Q_{\pm \mu \rho \sigma} =  e^{4 \varphi} [D_\mu( a_\rho a_\sigma) \mp
a_\rho (aB)_{\sigma \mu} \mp a_\sigma (aB)_{\rho \mu}]  \,  ,
\ee
where $(aB)_{\mu \nu} = a^\rho B_{\rho \mu \nu}$, $(aB)^2 = a^\rho B_{\rho \mu
\nu} a^\sigma B_\sigma{}^{\mu\nu}$ and $Da = D_\mu a^\mu$. First, we construct
the corresponding invariants
\be
{}^\star T^2_\pm \equiv {}^\star T_{\pm \mu \nu \rho} {}^\star T_\pm^{\mu \nu \rho} =
4 [B^2 - e^{4 \varphi}  (aB)^2]  \,  ,
\ee
\be
{}^\star Q_\pm^2 \equiv {}^\star Q_{\pm \mu \nu \rho} {}^\star Q_\pm^{\mu \nu \rho} =
8 (\partial \varphi)^2 +2 e^{4 \varphi} \left[(D_\mu a_\nu)  (D^\mu a^\nu)
+  (aB)^2 \right]  \,  ,
\ee
and
\be
{}^\star q^2 \equiv {}^\star q_\mu  {}^\star q^\mu  = {1 \over D^2} G^{\rho
\sigma} Q_{\pm \mu \rho \sigma} G^{\varepsilon \eta} Q_\pm{}^\mu
{}_{\varepsilon\eta} =
{16 \over D^2} (\partial \varphi)^2   \,  ,
\ee
where ${}^\star q_\mu$ is stringy Weyl vector defined in \eq{Wvs}. Note thet
all invariants are independent on $\pm$ indices and we put ${}^\star R_\pm =
{}^\star R$, ${}^\star T^2_\pm = {}^\star T^2$ and, ${}^\star Q^2_\pm =
{}^\star Q^2$.

We assume that Lagrangian is linear in these invariants and choose appropriate
coefficients in front of them
\be
{}^\star {\cal L} \equiv {}^\star R + {1 \over 48}
\left( 11 \,{}^\star T^2 -26 \,{}^\star Q^2 \right) + {1 \over 3}
\left( { 5 D \over 4} \right) {}^\star q^2  \, ,  \leq{lac}
\ee
in order to reproduce the expression  \eq{la}
\be
{}^\star {\cal L} = R -{1 \over 12} B^2 + 2 D^2 \varphi + {1 \over a^2}
\left[ 2 a^\mu \partial_\mu (Da) + a^\mu D_\mu (Da)
+ (Da)^2 - {13 \over 12} (D_\mu a_\nu)  (D^\mu a^\nu) \right]  \, .   \leq{lai}
\ee
If the condition \eq{lle} is satisfied, the Lagrangian \eq{lai}, up to the
term with factor ${1 \over a^2}$, coincides with that defined in \eq{la}.

The Lagrangian \eq{la} has been obtained from one-loop perturbative
computations. The higher loop corrections, generally depend on the
renormalization scheme (see \cite{Ts}). We argue that the term proportional to
${1 \over a^2}$ in ${}^\star {\cal L}$ originates from the higher orders
contribution. The reason is that there is difference between Lagrangian and
Hamiltonian perturbative approaches, \cite{BS1}. The leading order term of the
Hamiltonian contains $\Phi$ dependent part proportional to ${1 \over a^2}$,
while the leading order term of the Lagrangian is $\Phi$ independent. Because
the stringy invariants of the present paper are defined from Hamiltonian form
of field equations, \eq{lJ}-\eq{lF}, we expect that the term proportional to
${1 \over a^2}$ is a consequence of different perturbative approaches. Up to
this term, for $\varphi=\Phi$ we have ${}^\star {\cal L}= {\cal L}$.

\section{Conclusions}

In the present paper we showed that the probe string, as an extended object,
can see more space-time features then the probe particle -- torsion and
nonmetricity. We found their forms in terms of background fields, which defined
the target space geometry recognized by the string.

The equations of motion \eq{lJ}-\eq{lF}  help us to obtain the explicit
expression for stringy connection \eq{cdc}. It produces stringy torsion \eq{T}
and nonmetricity \eq{mp}, originated from the antisymmetric field $B_{\mu
\nu}$ and dilaton fields $\Phi$, respectively.

Let us clarify how space-time geometry depends on the background fields. In
the presence of the metric tensor $G_{\mu \nu}$, the space-time is of the
Riemann type. Inclusion of the antisymmetric field $B_{\mu \nu}$ produces the
Riemann-Cartan space-time. Appearance of the dilaton field $\Phi$ broke the
compatibility between metric tensor and stringy connection. When all three
background fields $G_{\mu \nu}$, $B_{\mu \nu}$ and $\Phi$ are present, the
string feels the complete stringy space-time.

Finally, we constructed the bosonic string space-time action in terms of
geometrical quantities. In order to find the integration measure invariant
under parallel transport we used the fact that the stringy Weyl vector is a
gradient of the scalar field $\varphi$. We also derive the Lagrangian as a
function of the stringy invariants: scalar curvature, torsion and
nonmetricity.

We discussed the connection between our result and that of the papers
\cite{CFMP}, in spite of their different origin. The standard result is
quantum and perturbative while our is classical and non perturbative. In
particular, our scalar field $\varphi$, defined in \eq{sf}, plays the role of
dilaton field $\Phi$, and has the same position in all expressions. Up to the
non-linear term proportional to ${1 \over a^2}$, (which is a consequence of a
different perturbatin theory in the Lagrangian and Hamiltonian approaches) for
$\varphi = \Phi$, these two actions are equal including the dilaton factor in
the integration measure.

It is well known that the dilaton dependent Weyl transformation
\be
G^E_{\mu \nu} = e^{-\fr{2(\Phi_0-\Phi)}{D-2}} G_{\mu \nu}
\ee
brings the Lagrangian to the Hilbert form
\be
S^E = \int dx \sqrt{-G^E} \left[ R -{1 \over 12} e^{- {8 \Phi \over D-2}} B^2 -
{ 4 \over D-2} (\partial \Phi)^2  \right]_E  \,  ,
\ee
where index $E$ means that all quantities are define in terms of the Einstein
metric, $G^E_{\mu \nu}$. In this form of the Lagrangian, the dilaton decouples
from the curvature, but it is still coupled to the torsion through the second
term.  As a consequence, neither of the two Lagrangians obeys the equivalence
principle, so that the change from the string frame to the Einstein one, does
not help us to choose a preferred definition of the metric (see second
reference \cite{GSW}).

Our approach prefers the so called string frame as a more fundamental, because
we offered clear geometrical interpretation  for it. In particular, the
preservation of the integration measure under parallel transport singles out
the form \eq{smf} for it. This is just characteristic of the string frame.

There is another reason in support of the string frame, \cite{GSW}. Only when
the action is written in terms of the fields originating from strings, the
constant part of the dilaton, $\Phi_0$, appears as an overall factor, as well
as the coupling constant in Yang-Mills theories.

Consequently, we show that string space-time action, in terms of geometrical
quantities, depends not only on curvature and torsion recognized by the probe
string, but also on nonmetricity, which causes absence of the equivalence
principle.

Let us mention one curiosity. It is known that the coefficient in front of the
Liouville action is proportional to the central charge and measures the
quantum braking of the classical symmetry. The contribution to the central
charge of the anticommuting ghosts $b, c$ corresponding to the conformal
symmetry and the commuting ghosts $\beta, \gamma$ corresponding to the
superconformal symmetry are ${-26 \over 48}$ and ${11 \over 48}$,
respectively. In definition of the Lagrangian ${}^\star {\cal L}$, \eq{lac},
the coefficients in front of the stringy nonmetricity and the stringy torsion
are just equal to the coefficients of the $b, c$ and $\beta, \gamma$ ghost
contributions. We do not find good reason to explain this similarity, but we
find interesting to mention this coincidence.

\appendix 

\section{World-sheet geometry}
\setcounter{equation}{0}

We used the notation of ref. \cite{BS1} expressing the intrinsic world-sheet metric tensor
$g_{\alpha \beta }$, in terms of the light-cone variables  $(h^+,h^-, F)$
\be
g_{\alpha \beta} =e^{2F} {\hat g}_{\alpha \beta}=
\fr{1}{2}e^{2F}\pmatrix{ -2h^-h^+    &  h^-+h^+ \cr
           h^-+h^+    &  -2      \cr }\,  .               \leq{g}
\ee
The world-sheet interval has a form
\be
ds^2 = g_{\alpha \beta} d \xi^\alpha d \xi^\beta = 2 d \xi^+ d
\xi^-  \,   ,
\ee
where
\be
d \xi^\pm = { \pm 1 \over \sqrt{2}} e^F ( d \xi^1 - h^\pm d \xi^0)
=  e^\pm{}_\alpha  d \xi^\alpha \,    .
\ee
The quantities $ e^\pm{}_\alpha$ define the light-cone one form basis,
$\theta^\pm = e^\pm{}_\alpha d \xi^\alpha$, and its inverse define the tangent
vector basis, $e_\pm = e_\pm{}^\alpha \partial_\alpha = \partial_\pm$.

In the tangent basis notation, the components of the arbitrary vector
$V_\alpha$ have the form
\be
V_{\pm} = e_{\pm}{}^\alpha V_\alpha =
{\sqrt{2} e^{-F} \over h^- -h^+} (V_0+h^{\mp}V_1)\,  .     \leq{vec}
\ee
The world-sheet covariant derivatives on tensor $X_n$ are
\be
\nabla_\pm X_n = (\pd_{\pm} +n \con_{\pm}) X_n   \,  ,        \leq{4.6}
\ee
where the number $n$ is sum of the indices, counting index $+$ with $1$ and
index $-$ with $-1$. The two dimensional covariant derivative $\nabla_\pm$ is
defined with respect to the connection

\be
\con_{\pm} =e^{-F}({\hat \con}_\pm \mp {\hat \partial}_\pm F) \,  , \qquad
{\hat \con}_\pm =\mp {\sqrt{2}\over h^- -h^+} h^{\mp \prime} \,  .
\ee

\end{document}